\documentclass[twocolumn,aps,prx]{revtex4}

\usepackage{graphics}
\usepackage{graphicx}
\usepackage{dcolumn}
\usepackage{bm}
\usepackage{amssymb,amsmath,graphicx,graphics,color,epstopdf}

\newcommand{\schw}{Schwarzschild }
\newcommand{\rn}{Reissner-Nordstr\"om }

\begin{document}


\title{On the Topological Nature of the Hawking Temperature of Black Holes}
\author{Charles W. Robson, Leone Di Mauro Villari and Fabio Biancalana}
\affiliation{School of Engineering and Physical Sciences, Heriot-Watt University, EH14 4AS Edinburgh, UK}
\date{\today}

\begin{abstract}

In this work we determine that the Hawking temperature of black holes possesses a purely topological nature. We find a very simple but powerful formula, based on a topological invariant known as the Euler characteristic, which is able to provide the exact Hawking temperature of any two-dimensional black hole -- and in fact of any metric that can be dimensionally reduced to two dimensions -- in any given coordinate system, introducing a covariant way to determine the temperature only using virtually trivial computations. We apply the topological temperature formula to several known black hole systems as well as to the Hawking emission of solitons of integrable equations.

\end{abstract}

\maketitle

\section{Introduction}

Black hole spacetimes are very special. It is known that they can be studied using topological methods, assigning an Euler characteristic $\chi$, a topological invariant, to each black hole \cite{Gibbons,Hawking,Eguchi,Liberati1}. By calculating the Euler characteristic of black hole spacetimes, several important features can be studied in a simple manner. One such feature is the black hole's entropy, which has been explored previously \cite{Liberati1,Teitelboim}, however we believe the use of $\chi$ to study black hole \emph{Hawking temperatures} has been neglected in the literature -- {although the topological nature of horizon temperatures has been previously highlighted by Padmanabhan using a formalism reminiscent of the Arahonov-Bohm effect\cite{Pad2}.} This is due to the fact that theorists generally study black holes in simple coordinate systems, for example the rotating black hole in the Kerr metric and a stationary black hole in Schwarzschild coordinates, where a temperature can be found easily using different methods adapted to each individual case \cite{Carroll1}. However, in many practical cases a black hole may be hidden within a complicated metric and highly nontrivial coordinate transformations required to unearth it. A topological formula for the Hawking temperature of a black hole related to $\chi$, working in any coordinate system, is therefore extremely useful. In this work we investigate this formula in depth for several different known black holes in two dimensions, exploring how the dependence of the metric and its curvature invariants on spacetime coordinates affects the formula.

Another novel feature of this work is that we use our new formula to investigate the temperature of \emph{solitons as black hole analogues}, building on our previous research into the new field of quantum soliton thermodynamics \cite{ourpaper}. This field explores the analogue Hawking radiation emission of soliton solutions of integrable equations. The reason why the topological formula for Hawking temperature is so important for the thermodynamics of solitonic systems is that these are described by induced metrics, derived from inverse scattering theory \cite{AKNS,Rogers}, of generally highly nontrivial form. These metrics are for the most part non-diagonal and have hidden symmetries and so a coordinate-independent route to the soliton's temperature is particularly apt. We believe that in future this topological approach will form a backbone for further research into quantum soliton thermodynamics, motivating the detailed treatment given here. \\

The plan of the paper is as follows. In Section \ref{formula} we introduce the main formula studied in this work, a new expression for the Hawking temperature of black hole systems in two dimensions. This formula we derive by investigating the features of a known topological invariant, the Euler characteristic $\chi$, the relevance of which to black hole spacetimes being described in Section \ref{math_back}. The applications of our Hawking temperature formula for several known black hole systems is then given in Section \ref{applic} after which the formula's notable advantage for studying analogue black holes, the central result of this paper, is demonstrated.

\section{A topological formula for the Hawking temperature of two-dimensional black holes} \label{formula}

Before introducing the formula's origins, we present it immediately in order to give the reader a flavour of its simplicity. The Hawking temperature of a two-dimensional (2d) black hole with time-independent Ricci scalar $R$ can be found as
\begin{equation} \label{eq:main1}
T_{\mathrm{H}}=\frac{\hbar c}{4\pi \chi k_{\mathrm{B}}}\sum_{j \leq \chi}\int_{r_{\mathrm{H_{j}}}}\sqrt{g}Rdr,
\end{equation}
where $T_{\rm H}$ is the Hawking temperature, $k_{\rm B}$ is the Boltzmann constant, $\hbar$ is the reduced Planck constant, $c$ is the speed of light in vacuum, $R(r)$ is the Ricci scalar which depends only on the `spatial' variable $r$, $g$ is the (Euclidean) metric determinant, $r_{\rm H_{j}}$ is the location of the $j$-th Killing horizon and $\chi$ is the Euler characteristic of the black hole's Euclidean `spacetime' (which counts its Killing horizons). {Of particular importance is the symbol $\sum_{j \leq \chi}$, which is intended to be a sum over all the Killing horizons, where one must pay attention to the sign of each term in the sum, which can be positive or negative depending on the specific features of each Killing horizon -- the overall result is however always positive. This subtle point will be explained in the discussions below. The single limit of integration signifies that the integral is to be evaluated only at the points $r_{H_{j}}$, the precise meaning of which will also be explained later.}

Eq. (\ref{eq:main1}) is the main new {\em covariant} formula studied in this paper and is derived from the known expression for the Euler characteristic $\chi$ of a black hole spacetime as we show below. This result highlights the {\em purely topological nature} of Hawking radiation. 

\section{Mathematical background} \label{math_back}

To make this paper self-contained we now review the derivation of the Euler characteristic of black holes.

The Euler characteristic $\chi$ is a remarkable topological invariant that provides important information on a manifold's structure. For a compact surface without boundary, the Euler characteristic simply counts its holes, giving its genus. For surfaces \emph{with boundary} (such as those we shall study below in the context of black hole spacetimes) the interpretation of $\chi$ is more subtle \cite{Morgan}.

The characteristic $\chi$ plays a fundamental role in the Gauss-Bonnet theorem, a key mathematical result which links geometry to topology. This theorem's generalisation to higher (even) dimensions, developed by Chern and other authors \cite{Allendoerfer,Chern1}, is the foundation upon which the application of $\chi$ to black hole physics is built, as will be elucidated below.

In $n$ (even) dimensions, the value of $\chi$ of a space can be defined as the integral over a certain density $G$ \cite{Morgan}:
\begin{equation} \label{eq:chi1}
\chi=\frac{2}{\mathrm{area}(S^{n})}\int_{\mathcal{M}^{n}} G,
\end{equation}
where $S^{n}$ is the $n$-sphere and $\mathrm{area}(S^{n})$ signifies its surface area, e.g. in 2d this area is $4\pi$. The density $G$ can be expressed in local coordinates as \cite{Morgan,Allendoerfer}
\begin{widetext}
\begin{equation} \label{eq:density}
G=\frac{1}{2^{n/2}n!g}R_{i_{1}i_{2}j_{1}j_{2}}R_{i_{3}i_{4}j_{3}j_{4}}...R_{i_{n-1}i_{n}j_{n-1}j_{n}}\epsilon^{i_{1}...i_{n}}\epsilon^{j_{1}...j_{n}},
\end{equation}
\end{widetext}
where $R_{\mu \nu \rho \tau}$ is the Riemann tensor defining the curvature of the space and $\epsilon^{ijkl}$ is the Levi-Civita symbol in four dimensions. In $n=2$ dimensions this density reduces to $G=R_{1212}/g=R/2$.

Two papers written by Chern \cite{Chern1,Chern2} during the Second World War, proving intrinsically (without any embeddings in higher dimensions) an extension of the Gauss-Bonnet theorem to $n$ even dimensions, is key for the application of topology to black holes, and so a brief summary of the key results of these papers will now be given, in the language of forms.

The Euler characteristic of an $n$-dimensional compact manifold $M^n$, which will later be identified with the black hole spacetime manifold capped at an outer boundary, can be defined as the integral of a density form $\Omega$ over the manifold by $\chi=\int_{M^n}\Omega$. This density $\Omega$ is necessarily given in Chern's papers in the guise of a differential form, however all calculations carried out later in this work employ the density in Riemannian coordinates, i.e. as defined in (\ref{eq:density}), because we wish to study black hole metrics in particular coordinate systems.

Chern proved that $\Omega$, originally defined in $M^n$, can also be defined in a larger manifold $M^{2n-1}$, itself formed by the unit vectors of $M^{n}$. The form $\Omega$ is equal to the exterior derivative of another density form, $\Pi$, of degree $n-1$ which is defined in $M^{2n-1}$ via $\Omega=-d\Pi$. Chern then showed that the integral of $\Omega$ over $M^n$ is equal to the same integral over a submanifold $V^n$ of $M^{2n-1}$ and by Stokes' theorem is also equal to the integral of $\Pi$ over the boundary of $V^n$.

Now, a manifold \emph{with boundary} requires an important correction to the value of $\chi$, which becomes \cite{Gibbons}:
\begin{equation} \label{eq:difference}
\chi=\int_{\partial V}\Pi-\int_{\partial M}\Pi.
\end{equation}
The submanifold $V^n$ of  $M^{2n-1}$ is crucial as its boundaries are defined to be the fixed points (the zeros) of the unit vector field defined in $M^n$. Strikingly, it is known that any unit vector field in $M^n$ can be chosen in order to find the value of $\chi$ for that manifold \cite{Chern2,Brass}.

A particularly elegant discussion of how fixed points of vector fields on manifold surfaces define the Euler characteristics of these surfaces, with relevance to black hole physics, is given in a work by Gibbons and Hawking using their concepts of ``nuts" and ``bolts"  \cite{Hawking}. These authors call isolated fixed points (that is, the zeros) of the Killing vector field on the manifold ``nuts'' and fixed point 2-surfaces they call ``bolts''. The Euler characteristic of the whole manifold can then be defined as the sum of the number of nuts and the Euler characteristic of each bolt. Each bolt, i.e. fixed point two-surface, must be compact in order for it to contribute to the total Euler characteristic, by definition, as $\chi$ is only defined for compact manifolds. Interestingly, if the Killing vector field only has nuts then the calculation of $\chi$ is identical to that using the Lefschetz fixed-point theorem for isometries \cite{Eguchi}.

Let us now see how the machinery defined above for a compact space can be utilised for black hole spacetimes. On first sight a spacetime would seem to be necessarily non-compact, spatially and temporally. We already mentioned that the boundaries of $V^n$ are the fixed points of the unit vector field defined in $M^n$. A choice that simplifies results substantially is to define this unit vector field to be a \emph{Killing vector field} of the black hole spacetime. For a 2d Schwarzschild black hole the Killing vector field is associated to time symmetry and the field vanishes only at a Killing horizon, therefore $V^n$ has a boundary associated to the Killing horizon (which for a Schwarzschild black hole coincides with its well-known unique event horizon). We have now defined one of the boundaries of $V^n$ to be at the Killing horizon of the black hole, which we will denote $r_{H}$. Does it have any other boundaries? It has one other, at the boundary of the original manifold $M^n$, which we call $r_{0}$.

What about $M^n$? This is the black hole spacetime capped at $r_{0}$, which defines its only boundary. Taking all these results into account (\ref{eq:difference}) becomes:
\begin{equation}
\chi=\int_{r_{0}}\Pi-\int_{r_{\mathrm{H}}}\Pi-\int_{r_{0}}\Pi=-\int_{r_{\mathrm{H}}}\Pi.
\end{equation}
This result means that the \emph{particular choice} of the manifold's outer boundary position (although the existence of the boundary is key for making the spacetime compact) is irrelevant for the calculation of the Euler characteristic of the black hole spacetime. Put briefly: the outer boundary always cancels out in the calculation of the Euler characteristic. The integral should therefore only be evaluated at the Killing horizon, hence the integral in our main formula Eq. (\ref{eq:main1}) contains only the lower limit of integration.

Now we see that the integral formula (\ref{eq:chi1}) defining $\chi$ should only be evaluated at one limit, namely at the Killing horizon of the spacetime being studied. If a spacetime has more than one Killing horizon then each must be taken into account separately, as we show in the next section.

In 2d for $\chi=1$, the density formula (\ref{eq:density}) leads to an Euler characteristic of:
\begin{equation}
\int d^2 x \sqrt{g} \frac{R}{4\pi}=1,
\end{equation}
Two-dimensional Schwarzschild black holes are known to have topology $\mathbb{R}^2$ and thus satisfy $\chi=1$ \cite{Gibbons}. The temperature of these black holes emerges from the above expression.

Applying a Wick rotation to the spacetime being studied makes the time dimension compact and this rotation to Euclidean space, setting the period of Euclidean time to be inverse temperature $\beta$, provides the appropriate limits for the Euclidean time integral \cite{Gibbons,Pad}. The remaining spatial integral has only one limit at which to be evaluated, at the Killing horizon, as already explained.

The expression for $\chi$ thus becomes: $\int_{0}^{\beta}dt \int_{r_{\mathrm{H}}} dr \sqrt{g}(R/4\pi)=1$. If $R$ is time-independent {(where the meaning of the `time' variable of course depends on the specific coordinate system adopted)}, then the expression, after units have been introduced, becomes:
\begin{equation} \label{eq:Fabios_eq}
\frac{\hbar c}{4\pi k_{\mathrm{B}}T_{\mathrm{H}}}\int_{r_{\mathrm{H}}}\sqrt{g}Rdr=1,
\end{equation}
i.e. the expression presented earlier (\ref{eq:main1}) for the case of a black hole with one Killing horizon.

With all of the reasoning above we have shown, for the first time to our knowledge, that {\em the Hawking temperature is a purely topological quantity}, explaining the covariant form of the equation and its known applicability in any coordinate system, as demonstrated in a few cases below.

\section{Applications} \label{applic}

\subsection{Black holes}

\subsubsection{Static black holes}

Let us now apply the temperature formula (\ref{eq:main1}) to several canonical cases of black hole systems, starting with that of the two-dimensional static, uncharged black hole.

In Euclidean Schwarzschild coordinates the static black hole's metric is:
\begin{equation} \label{eq:Schwarz2d}
ds^2 = \left( 1-\frac{2GM}{c^2 r} \right)d\tau^2 + \left( 1-\frac{2GM}{c^2 r} \right)^{-1 }dr^2.
\end{equation}
The Ricci scalar for this spacetime is $R=(4GM/c^2 r^3)$ and clearly $g=1$, therefore (\ref{eq:main1}) gives:
\begin{equation}
T_{\mathrm{H}}=\frac{\hbar c^3}{8\pi GM k_{\mathrm{B}}},
\end{equation}
the known static, uncharged black hole temperature. $G$ and $M$ are Newton's constant and the black hole's mass respectively, $c$ and $k_{\mathrm{B}}$ are the speed of the light in vacuum and Boltzmann's constant.

An arbitrary coordinate transformation will lead to a different metric describing the static black hole however, crucially, due to the topological nature of the temperature formula, the same static black hole temperature will emerge from any coordinate system, as can be readily checked.

The appropriate method to find the horizon position of a black hole in an arbitrary coordinate system is either to find where the Killing vectors of the metric vanish or, usually more simply, to use an invariant known as the Karlhede scalar, defined as $K_{2}\equiv R^{\mu\nu\rho\sigma;\tau}R_{\mu\nu\rho\sigma;\tau}$, where the symbol ``;" indicates a covariant derivative \cite{Chandra,Carroll1}. This invariant vanishes at the event horizon of \emph{most} black holes \cite{karlhede_scal}. The roots of the Karlhede scalar are known to provide a means to detect the position of the event horizon of a spherically symmetric black hole (for instance a \schw or a \rn black hole), where $K_{2}$ vanishes and changes sign after traversing the horizon. However, for some types of black hole, $K_{2}$ is known to possess extra roots that are associated to other liminal regions -- for example in the 4d Kerr metric, which describes uncharged rotating black holes, $K_{2}$ vanishes at the ergosphere \cite{karlhede_scal}.

In the next section we shall look at a 2d description of rotating black holes.

\subsubsection{Rotating black holes}

In this and the following sections all physical constants are set to unity to simplify the formulae and metrics, without any loss of generality.

A black hole with angular momentum is generally studied in 3+1 dimensions (described by the so-called Kerr metric), however recently an effective 1+1-dimensional metric for the system was derived \cite{Murata}. It has the form:
\begin{equation} \label{eq:rotating_bh}
ds^2=-f(r)dt^2 + \frac{1}{f(r)}dr^2,
\end{equation}
where $f(r)=\left( r^2 - 2Mr + a^2 \right)/(r_{+}^2 + a^2)$, with the black hole's angular momentum per unit mass denoted by $a$, its mass as usual by $M$, and its outer horizon by $r_{+}=M+\sqrt{M^2-a^2}$.

A Wick rotation transforms metric (\ref{eq:rotating_bh}), via $\tau=it$, to $ds^2=f(r)d\tau^2 + f(r)^{-1}dr^2$, with the same definition of $f(r)$ as above. It has been argued that a rotating black hole's angular momentum parameter $a$ should be left unaltered after a Wick rotation in order to properly define its Euclidean metric and our calculations below support this view \cite{Brown1}.

We see that $f(r)$ has two roots $r_{\pm}=M \pm \sqrt{M^2-a^2}$ and hence two fixed points of its Killing vector field associated to time isometry, therefore $\chi=2$. The fact that the Killing vector field has two zeros adds another boundary to manifold $V^{n}$ defined in Section \ref{math_back}, making two integrals necessary for defining $\chi$ and hence $T_{\mathrm{H}}$.

Therefore the form of $T_{\mathrm{H}}$ becomes (note that $g=1$):
\begin{equation} \label{eq:rotating_temp}
T_{\mathrm{H}}=\frac{1}{2}\left( \frac{1}{4\pi}\int_{r_{+}}Rdr -  \frac{1}{4\pi}\int_{r_{-}}Rdr \right).
\end{equation}
For the metric (\ref{eq:rotating_bh}) the formula (\ref{eq:rotating_temp}) gives the known rotating black hole temperature of \cite{Murata}:
\begin{equation}
T_{\mathrm{H}}=\frac{\sqrt{M^2-a^2}}{4\pi M \left( M+\sqrt{M^2-a^2} \right)},
\end{equation}
demonstrating again the analytical power of our general formula (\ref{eq:main1}).

\subsubsection{Charged black holes}

A black hole with electrical charge $Q$ can be described by the Reissner-Nordstr{\" o}m metric, which with Euclidean signature and after a dimensional reduction to two dimensions is \cite{Bobev}:
\begin{equation} \label{eq:charged_metric}
ds^2=\left( 1-\frac{r_\mathrm{s}}{r}-\frac{r_{\mathrm{Q}}^2}{r^2} \right)d\tau^2 + \left( 1-\frac{r_\mathrm{s}}{r}-\frac{r_{\mathrm{Q}}^2}{r^2} \right)^{-1} dr^2,
\end{equation}
where $r_{\mathrm{s}}=2M$ and $r_{\mathrm{Q}}^2=Q^2$ are length scales of the system associated to mass and charge. Note that (\ref{eq:charged_metric}) is related to the Lorentzian signature metric by both a Wick-rotated time and charge.

Metric (\ref{eq:charged_metric}) has one Killing horizon associated to time isometry located at $r_{+}=\frac{1}{2}\left( r_{\mathrm{s}} + \sqrt{ r_{\mathrm{s}}^2 + 4  r_{\mathrm{Q}}^2} \right)$. The only positive root of the Karlhede scalar $K_{2}$ is $r_{+}$ when applied to the charged black hole metric (\ref{eq:charged_metric}), pinpointing its horizon position.

Once this geometry is inputted into (\ref{eq:main1}) and one Wick rotates back, the standard Reissner-Nordstr{\" o}m black hole temperature is found:
\begin{equation}
T_{\mathrm{H}}=\frac{-4r_{\mathrm{Q}}^2 + r_{\mathrm{s}}\left( r_{\mathrm{s}} + \sqrt{r_{\mathrm{s}}^2 - 4 r_{\mathrm{Q}}^2} \right)}{\pi \left( r_{\mathrm{s}} + \sqrt{r_{\mathrm{s}}^2 - 4 r_{\mathrm{Q}}^2} \right)^3}.
\end{equation}

It may seem that the treatment given in this work is somewhat limited as it covers only the properties of two-dimensional black holes, however this is not the case {for two reasons: one, lower-dimensional black holes are interesting in their own right as they can be studied more easily than their four-dimensional counterparts and are used to attack conceptual problems in quantum gravity \cite{Carlip}; and two, it is known that higher-dimensional black hole metrics can in many cases be \emph{dimensionally reduced} to 2d. For example, scalar fields on the standard four-dimensional Schwarzschild, \rn and Kerr metric backgrounds can be described by field theory on effective 2d metrics. This is done by keeping only near-horizon dominant terms in the action describing scalar fields on these backgrounds \cite{Murata}. In such a way higher-dimensional black holes can be described by 2d effective metrics amenable to our Hawking temperature formula (\ref{eq:main1}). It should be noted that there are other ways to dimensionally reduce higher-dimensional gravity theory which have associated subtleties, involving the dynamics of dilatonic fields, but this is beyond the scope of the current work. For a treatment of these subtleties see \cite{Vass}.}

\subsubsection{AdS black holes}

Anti-de Sitter space has a geometry of constant negative curvature and plays a key role in the AdS/CFT correspondence and the field of holography, two rich areas of modern physics \cite{Susskind}.

A black hole in 2d anti-de Sitter ($\mathrm{AdS}_2$) space has a Euclidean metric description \cite{Hubeny}:
\begin{widetext}
\begin{equation}
ds^2=\left( 1+\frac{r^2}{l^2}-\frac{r}{r_{+}}\left( \frac{r_{+}^2}{l^2}+1 \right) \right)d\tau^2 + \left( 1+\frac{r^2}{l^2}-\frac{r}{r_{+}}\left( \frac{r_{+}^2}{l^2}+1 \right) \right)^{-1} dr^2,
\end{equation}
\end{widetext}
which has two Killing horizons defined by time symmetry positioned at $r_{-}=l^2/r_{+}$ and $r_{+}$. The AdS length scale is defined by $l$ \cite{Hubeny}. The metric above is related to the Lorentzian signature version only by a Wick-rotated time. As it has two Killing horizons, the Euler characteristic is two, as is known from topological considerations \cite{Charmousis}, and the form of (\ref{eq:main1}) with two integrals as in the rotating black hole case must be used. 

Evaluating (\ref{eq:main1}) after taking both horizons into account then gives the known $\mathrm{AdS}_2$ black hole Hawking temperature of \cite{Hubeny}:
\begin{equation}
T_{\mathrm{H}}=\frac{r_{+}^2-l^2}{4 \pi r_{+}l^2}.
\end{equation}

\subsection{Solitons as analogue black holes}

So far we have only used the Euler characteristic-based Hawking temperature formula (\ref{eq:main1}) to verify known astrophysical black hole temperatures. This was to show its effectiveness in different situations and its complete coordinate independence. For example, its efficacy for black holes with differing numbers of horizons and with differing amounts of ``hair", i.e. externally measurable parameters \cite{Thorne}. The main focus of this paper however is in the power of this method for \emph{black hole analogues}. Hawking radiation from astrophysical black holes is so weak that its experimental verification is, at least for the foreseeable future, completely impractical. Therefore in recent years a large amount of effort in the community has been invested in measuring the analogue of Hawking radiation in other, laboratory-based, settings including hydrodynamical, optical and cold-atom systems \cite{Faccio,Ulf,Jeff,Drori}.

The present authors have previously studied the possibility of detecting the analogue of Hawking radiation from solitons, stable solutions of integrable equations, finding a thermodynamical first law for each \cite{ourpaper}. A prediction of a half-life of an optical soliton propagating in an optical fiber due to this emission was made. This study of quantum soliton thermodynamics was based upon an induced metric, derivable for any two-dimensional soliton of an integrable nonlinear evolution equation. It is known that any two-dimensional metric can be put into Schwarzschild-like form and thus there is a {\em complete formal equivalence} between these solitons and black holes \cite{Chandra}. Whether these solitons' induced metrics have real-valued Hawking temperatures and horizons, and therefore emit radiation, must be investigated by studying each metric's properties. However, each 2d soliton can potentially Hawking radiate. Several highly nontrivial coordinate transformations are generally required to cast a soliton's metric in Schwarzschild-like form and thence to find its associated temperature. Rather than performing the previously obligatory series of involved coordinate transformations, we believe that the topological formula described in the present work is much more suitable, as we demonstrate below. For more information on the construction of soliton metrics please see \cite{ourpaper,Rogers}.

We focus now on the richest case of quantum soliton thermodynamics: the soliton solution to the nonlinear Schr{\" o}dinger (NLS) equation. This equation appears in many places in physics, from fiber optics to Bose-Einstein condensates \cite{Agrawal,Bose}. The NLS equation itself is given by
\begin{equation}
iu_{t}+u_{xx}+2|u|^2 u=0
\end{equation}
where $t$ and $x$ are time and space respectively and $u$ is a complex scalar field. The NLS equation has an associated metric with line element \cite{ourpaper}:
\begin{equation}
ds^2=4dx^2 + 8vdxdt + \left( 16|u|^2 + 4v^2 -16B^2 \right) dt^2,
\end{equation}
where $v$ and $B$ are the soliton's velocity and amplitude respectively. The soliton solution $|u|=B\mathrm{sech}(Bx)$ is to be inputted into the line element above in order to define the NLS soliton's induced metric.

The soliton metric as it is would have a Lorentzian signature and so must be Wick-rotated to a Euclidean signature as explained in Section \ref{math_back}. This is done by applying $t \rightarrow i\tau$ and $v \rightarrow -iv$ after which the line element becomes $ds^2=4dx^2 + 8vdxd\tau - \left( 16|u|^2 - 4v^2 -16B^2 \right) d\tau^2$. Now, it is known that all two-dimensional metrics can be put into Schwarzschild-like form, but how does one find the horizon position in a general coordinate system? To answer this question, we note that the Euclidean NLS metric written above has a Karlhede scalar of $K_{2}=\frac{B^4}{4}\mathrm{sech}^4 (Bx)\left( 4B^2 + v^2 - 4B^2\mathrm{sech}^2 (Bx) \right)$ which vanishes at the horizon position $x_{\mathrm{H}}=\frac{1}{B}\mathrm{sech}^{-1}\left( \sqrt{1+\frac{v^2}{4B^2}} \right)$. Evaluating (\ref{eq:main1}) for the Euclidean NLS metric (and afterwards Wick rotating the velocity back) produces the known NLS soliton Hawking temperature \cite{ourpaper}:
\begin{equation}
T_{\mathrm{H}}=\frac{B^2}{\pi}\left( 1-\frac{v^2}{4B^2} \right).
\end{equation}
A basic study of the NLS soliton's Hamiltonian combined with the above temperature then reveals the soliton's entropy to be $S=2\pi B$ \cite{ourpaper}.

Another example of a solitonic black hole analogue comes from the sine-Gordon equation $u_{\xi \xi}+u_{\tau \tau}=\mathrm{sin}(u)$, which has an interesting solution, known as a kink, moving with velocity $v$ \cite{Leone1,Gegenberg}. This solution is described by $u=4\mathrm{tan}^{-1}\left( e^{\gamma (\xi - v\tau)} \right)$, where $\gamma \equiv \left( 1+v^2 \right)^{-1/2}$.

The kink has an induced Euclidean metric:
\begin{equation}
ds^2=-\left( v^2 - r^2 \right)d\tau^2 -\left( v^2 - r^2 \right)^{-1}dr^2,
\end{equation}
with a horizon position that can without difficulty be read off from the line element and is located at $r=v$. The topological formula (\ref{eq:main1}) then gives:
\begin{equation}
T_{\mathrm{H}}=\frac{v}{2\pi},
\end{equation} \\
the known sine-Gordon kink Hawking temperature \cite{Leone1}.

Any 2d integrable soliton solution can be studied as an analogue of a black hole following the procedure given here and in our previous work \cite{ourpaper}. Using these methods, the KdV equation's soliton solution has been shown not to Hawking radiate due to a complex-valued event horizon, however we hope that the quantum thermodynamics of other solitons can be probed in the future, uncovering other examples of quantum emission. We have shown in this work that a topological method based on the Euler characteristic $\chi$ is particularly well suited for studying black hole analogues, further unveiling the profound link between solitons and black holes.

\section{Conclusions}

In this work we have presented a formula Eq. (\ref{eq:main1}), derived by studying the Euler characteristic of black hole spacetimes, which is able to provide a simple way of calculating the Hawking temperatures of black holes in arbitrary coordinate systems, a topological application we believe to have been neglected so far in the literature. We have shown its efficacy for several important black hole systems, with varying numbers of parameters and Killing horizons. It is argued that the most powerful application of this topological method is to the new field of quantum soliton thermodynamics. This new field, describing solitons of integrable equations as black hole analogues, should benefit greatly from the topological methods presented here as, in general, the thermodynamics of these solitons are encoded in highly nontrivial metrics which can easily be resolved using topology.

\section*{Acknowledgments}
This research was partially funded by the German Max Planck Society for the Advancement of Science (MPG). LDMV acknowledges support from EPSRC through CM-CDT.

\end{document}